\documentclass{PoS}

\usepackage{epsfig}
\usepackage{graphics,color}

\title{Two topics from lattice NRQCD at non-zero temperature: heavy quark mass dependence and S-wave bottomonium states moving in a thermal bath}

\ShortTitle{lattice NRQCD at non-zero T}

\author{\speaker{Seyong Kim} \\
        Department of Physics, Sejong University, Seoul 143-747, Korea \\
        E-mail: \email{skim@sejong.ac.kr}}

\author{Gert Aarts, Chris Allton \\
        Department of Physics, Swansea University, Swansea, United Kingdom}

\author{Maria Paola Lombardo \\
        INFN-Laboratori Nazionali di Frascati, I-00044, Frascati (RM), Italy}

\author{Mehmet B. Oktay \\
        Physics Department, University of Utah, Salt Lake City, Utah, USA}

\author{Sinead M. Ryan \\
        School of Mathematics, Trinity College, Dublin 2, Ireland}

\author{Donald K. Sinclair \\
        HEP Division, Argonne National Laboratory, 9700 South Cass Avenue, Argonne, Ill 60439, USA}

\author{Jon-Ivar Skullerud \\
        Department of Mathematical Physics, National University of Ireland Maynooth, Maynooth, Ireland}

\abstract{Using Non-Relativistic QCD (NRQCD), we study heavy quark
  mass dependence of S-wave and P-wave bottomonium correlators for
  $0.42 T_c <= T <= 2.09 T_c$ and study spectral functions of S-wave
  bottomonium states moving in a thermal bath at these temperatures
  using Maximum Entropy Method with NRQCD kernel. For the studied
  momentum range, the energy of moving states shows quadratic
  momentum-dependence and the width of moving states does not show
  significant changes as the momentum of bottomonium is
  increased. Also, we find that in correlator ratios, the temperature
  effect is larger than the effect caused by $20\%$ change in the
  bottom quark mass.
}

\FullConference{The 30th International Symposium on Lattice Field Theory - Lattice 2012,\\
		June 24-29, 2012\\
		Cairns, Australia}

\begin{document}

\section{Introduction}

Quark-Gluon-Plasma (QGP) is a state of matter in the Strong
Interaction system which is actively sought through theoretical study
and heavy ion collision at high energy (or high temperature) and at
low energy (or high baryon density) and must have played an important
role in the evolution of early universe
\cite{Yagi:2005yb}. Unfortunately at physical quark masses, chiral
symmetry restoration/deconfinement transition in QCD from normal
hadronic matter to QGP matter is a cross-over transition and lacks
characteristic behaviors of order parameter in a first order or a
second order thermodynamic phase transition across the thermodynamic
transition. Thus, finding and understanding physical phenomena which
can serve as signatures for QGP matter formation is quite crucial.

Physics of heavy quarkonium in medium may be one of such candidate
phenomena \cite{Matsui:1986dk}. In contrast to large uncertainty in
our understanding of light hadron production and decays, inclusive
productions and decays of heavy quarkonium in proton--proton or
proton--anti-proton collision, which are required for the comparison
with those in heavy ion collisions, is well understood:
Non-Relativistic QCD (NRQCD) factorization in which the scale above
heavy quark (like charm quark or bottom quark) mass is ``integrated
away'' and the smallness of heavy quark velocity at the quarkonium
rest frame offers ``small parameter'' is applicable in heavy
quarkonium \cite{Bodwin:1994jh}. Short distance effects can be
calculated via perturbative expansion in $\alpha_s$ and long distance
non-perturbative matrix elements can be computed either using lattice
NRQCD at zero
temperature\cite{Bodwin:1996tg,Bodwin:2001mk,Bodwin:2005gg} or can be
fitted using experimental data (see e.g.,
\cite{Butenschoen:2010rq}). Based on our understanding of quarkonium
physics in p--p collision, medium effect for quarkonium in heavy ion
collision may be better disentangled. In particular, bottomonium
system offers a better probe for QGP formation since there are clearer
seperations between the long distance physics and the short distance
physics, compared to those in charmonium system due to smaller quark
velocity ($v_b^2 \sim 0.1$ compared to $v_c^2 \sim 0.3$ for charmonium
system) at the quarkonium rest frame.

In recent years, we studied the behavior of S-wave and P-wave
bottomonium states for a temperature range $0.4 T_c \le T \le
2.1 T_c$ using NRQCD in non-zero temperature on anisotropic lattices
($12^3 \times N_t$) with two light quark flavors
\cite{Aarts:2010ek,Aarts:2011sm}. Details on how dynamical gauge field
is generated and what the simulation parameters are can be found in
\cite{Aarts:2010ek,Aarts:2011sm,Morrin:2006tf}. Our calculation method
is explained in \cite{Aarts:2010ek,Aarts:2011sm}. We expect that
in-medium effect such as temperature ($T$) effect or baryon chemical
potential ($\mu$) effect can be accommodated by NRQCD as long as
$\frac{T}{M} \ll 1$ or $\frac{\mu}{M}\ll 1$ where $M$ is the heavy
quark mass (see, e.g. \cite{Hands:2012yy} for baryon chemical
potential efffect in NRQC$_2$D). Indeed our study
\cite{Aarts:2010ek,Aarts:2011sm} meets such expectation. We found that
P-wave bottomonium states ($\chi_{bJ}$) melt immediately above $T_c$,
and bound state signals for S-wave channel ($\eta_b, \Upsilon$)
survive above $T_c$ \cite{Aarts:2010ek}. Closer inspection of the
spectral functions for S-wave correlators ($\chi_b, \Upsilon$), which
is obtained by the Maximum Entropy Method (MEM) with NRQCD kernel,
shows that $(1S)$ peaks appears to survive and the excited state peaks
are suppressed \cite{Aarts:2011sm}. This observation is consistent
with recent CMS experiment which shows sequential suppression of
$(3S), (2S)$ $\Upsilon$ states and survival of $(1S) \Upsilon$
\cite{Chatrchyan:2011pe,Chatrchyan:2012fr}.

\begin{figure}[t]
\begin{center}
\epsfig{figure=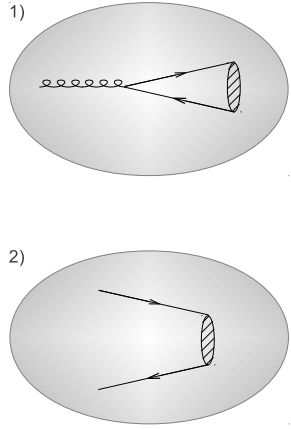,width=0.3\textwidth}
\end{center}
\caption{Quarkonium production processes: (a)fragmentation from
  gluon (top), (b) recombination of thermalized heavy quark--heavy
  anti-quark (bottom)}
 \label{fig:1}
\end{figure}

Here, we discuss further lattice NRQCD study of ours on bottomonium at
non-zero temperature. First, bottomonium with non-zero momentum is
studied (for full report, see \cite{Aarts:2012ka}). Quarkonium
production in hadron collision proceeds: parton evolution from hadron,
then partonic scattering process (gluon fusion, heavy quark
recombination, gluon Compton scattering, and light quark--light quark
scattering). Among these partonic processes, quarkonium production is
mostly dominated by gluon fusion. Gluon subsequently fragments into
quarkonium. In this case, fragmenting gluon will have large energy and
momentum. If quarkonium is produced in thermal bath, then it will be
moving relative to thermal bath. In contrast, if quarkonium is
produced through recombination of thermalized heavy quark and heavy
anti-quark (albeit less probable), produced quarkonium is at rest in
thermal bath. Schematic view on these two processes is shown in
Fig. \ref{fig:1}. Since a thermal bath defines a preferred rest frame,
``moving'' effect will induce $\frac{1}{2} M v^2$ shift in the mass
compared to the mass of quarkonium produced at rest. This shift may be
larger than experimental resolution for muon detection and may be
observable.

Secondly, we start to investigate systematic errors associated with
our computation in \cite{Aarts:2010ek,Aarts:2011sm}. As a first step,
we investigate how $10 \sim 20 \%$ change in bottom quark mass affects
previous observations made in \cite{Aarts:2010ek,Aarts:2011sm} by
calculating bottomonium correlators with $M a_s = 3.7, 4.14$ and $5.4$
and looking at the correlator ratios.

\section{S-wave bottomonium moving in thermal bath}

\begin{figure}[t]
\begin{center}
\epsfig{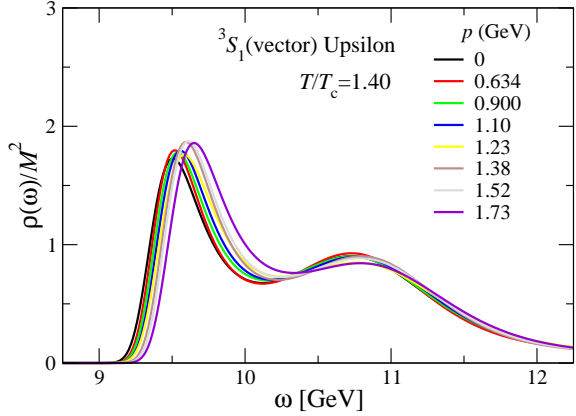}
\end{center}
\caption{$\Upsilon$ spectral functions on $12^3 \times 24$ for various
  momenta} 
 \label{fig:2}
\end{figure}

\begin{figure}[t]
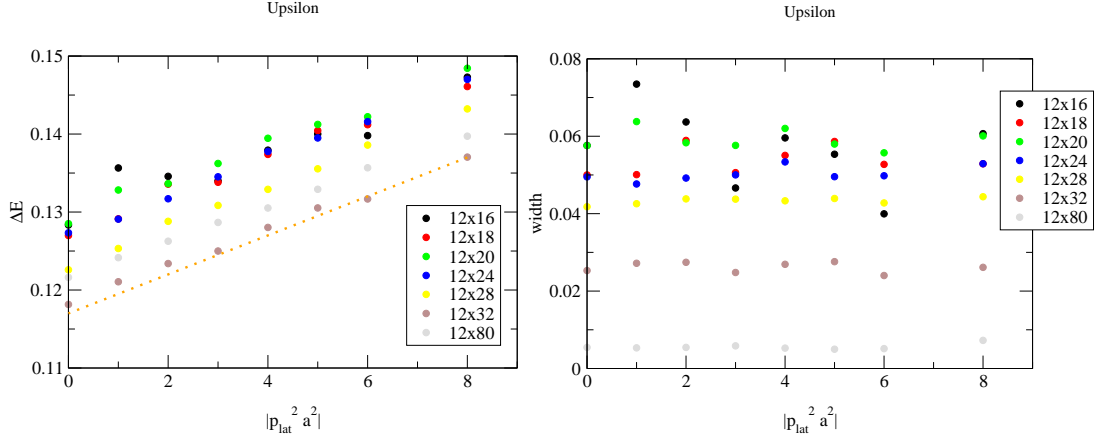

\begin{center}
\epsfig{figure=upsilon_mass_vs_p_2_all_Nt.eps,width=0.45\textwidth}
\epsfig{figure=upsilon_width_vs_p_2_all_Nt.eps,width=0.5\textwidth}
\end{center}
\caption{$\Upsilon (1S)$ energy vs. various momenta from the first
  peak of the spectral functions at each $T/T_c = 2.09, 1.86, 1.68,
  1.40, 1.20, 1.05,$ and $0.42$ for $N_t = 16, 18, 20, 24, 28, 32$,
  $80$ respectively (left). $\Upsilon (1S)$ width vs. various momenta
  from the width of the first peak in spectral functions
  (right)} \label{fig:3} 
\end{figure}

We consider $\eta_b$ and $\Upsilon$ which are moving with regard to
the thermal bath, i.e., S-wave bottomonium states with
momentum. Non-zero momentum quarkonium state is obtained by injecting
momentum to the source for NRQCD quark propagator computation and
projecting out the injected momentum component from the quarkonium
propagator which is obtained by combining the zero-momentum NRQCD
quark propagator with the momentum injected NRQCD quark
propagator. Momenta, $\vec{k} = \frac{2\pi}{L} (n_1,n_2,n_3)$ where
$\vec{n} = (0,0,0), (1,0,0), (1,1,0), (1,1,1), (2,0,0), (2,1,0)$, and
$(2,2,0)$ with $L=12$, have been considered. Since $M a_s = 4.5$ is
used for the bottom quark mass in NRQCD propagator computation and
$M_{\eta(1S)} = 9.391$ GeV and $M_{\Upsilon (1S)} = 9.460$ GeV
\cite{Beringer:1900zz}, these quarkonium momentum range amount to
quarkonium velocity, $0.07 \stackrel{<}{\sim} \;\; v \;\;
\stackrel{<}{\sim} 0.18$ with $\vec{p}^2 a_s^2 = 4 \sum_{i=1,3} \sin^2
\frac{k_i}{2}$.  From the correlators, spectral functions defined as
\begin{equation}
G(\tau, \vec{p}) = \int \frac{d \omega}{2\pi} K(\tau, \omega) \rho
(\omega, \vec{p}), \;\;\;\;\; K(\tau, \omega) = e^{-\omega \tau}
\end{equation}
are computed using MEM. Fig. \ref{fig:2} shows a typical spectral
function as a function of $|\vec{p}|$. For each momentum, the spectral
function exhibits a distinct shape. From this, the peak and width can
be extracted and Fig. \ref{fig:3} show extracted peak values and width
values which corresponds the $\Upsilon (1S)$ state. To a leading order
in $\vec{p}^2$ and $T$, we see that the $1S$ peak roughly follows
\begin{equation}
\Delta E (\vec{p}, T) \simeq \Delta E (\vec{p} = 0, T = 0) +
\frac{\vec{p}^2}{2M} + c (\alpha_s) T \label{eqDeltaE}
\end{equation}
In general, continuum NRQCD dispersion is
\begin{equation}
E = M_0 + \frac{\vec{p}^2}{2M_2} - \frac{(\vec{p}^2)}{8 M_4^3} + \cdots 
\end{equation}
Since we use ${\cal O} (v^4)$ NRQCD lagrangian to compute bottomonium
propagator , $M_0$ has ${\cal O} (v^4)$ error, $M_2$ has ${\cal O}
(v^2)$ error, and $M_4$ ${\cal O} (v^0)$ respectively. Using the form
of $M_0 + 4\sum_i( \sin (k_i/2)^2)/2M_2 -
(4\sum_i(\sin(k_i/2)^2)^2)/8M_2^3$ to fit $N_t = 80$ $\Upsilon$
correlator, we get 11.05 GeV for the kinetic mass ($M_2$) of $\Upsilon
(1S)$ state with $M a_s = 4.5$. The fitting form $M_0 + 4\sum_i(\sin
(k_i/2)^2)/2M_2$ gives 11.12 GeV for $M_2$. Thus Eq. (\ref{eqDeltaE})
is suitable to describe our lattice result. On the other hand, the
$1S$ width is independent of $\vec{p}^2$ and is just proportional to
$T$. 

This lattice NRQCD finding can be compared with effective theory
prediction. NRQCD bound state at rest in thermal bath below the
melting temperature has been considered in \cite{Brambilla:2010vq} and
non-relativistic bound state which is moving with regard to thermal
bath has been considered in QED (in the rest frame of the bound state,
i.e., bound state in moving thermal bath)
\cite{Escobedo:2011ie}. Finding in \cite{Escobedo:2011ie} for the
width can be summarized as 
\begin{equation}
\frac{\Gamma_v}{\Gamma_0} = \frac{\sqrt{1-v^2}}{2v} \log
\left(\frac{1+v}{1-v} \right) \rightarrow 1 - \frac{2}{3} v^2 + {\cal
  O} (v^4)
\end{equation}
where $\Gamma_0$ is the width of the bound state not moving with
regard to thermal bath and $\Gamma_v$ is the width of the bound state
moving with the velocity $v$ in thermal bath. With $v^2 < 0.04$,
``moving effect'' in the width is beyond the precision level of our
computation, which explains negligible momentum dependence of the
width in the right figure of Fig. \ref{fig:3}.

\section{Mass dependence of bottomonium correlators}

Our previous bottomonium study was done with bottom quark mass, $M a_s
= 4.5$. The same NRQCD correlator calculation is repeated with $M a_s
= 3.7 (-20 \%), 4.14 (-10 \%)$ and $5.4 (+ 20\%)$ in order to
investigate how small change in bottom quark mass affects bottomonium
spectrum and how important this effect is compared to non-zero
temperature effect. Since lattice NRQCD is an effective theory and is
consistent when $M a_\tau \sim 1$, all these choices ($M a_\tau =
0.62, 0.75, 0.9$ for $M a_s = 3,7, 4.5, 5.4$ respectively) are
expected to be valid.

\begin{figure}[t]
\begin{center}
\epsfig{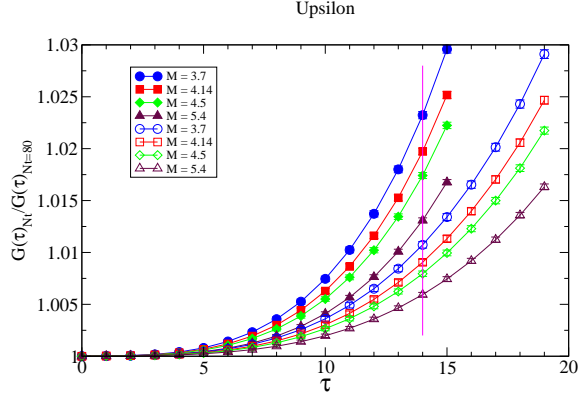}
\end{center}
\caption{$\Upsilon$ correlator ratio of $G(\tau;N_t)$ to
  $G(\tau;{N_t = 80})$ for $N_t = 16$ (filled symbol)
  and $18$ (open symbol)}
 \label{fig:5}
\end{figure}

Lattice NRQCD quarkonium correlator behaves as $A e^{-\Delta E \tau}$
when a bound state is formed and fitting $\Upsilon$ correlator
calculated on $N_t = 80$ configurations gives $\Delta E_\Upsilon (M,T
\sim 0)$. Using experimental value for $M_\Upsilon$, we determine the
unknown constant for NRQCD spectrum by
\begin{equation}
M_\Upsilon^{\rm exp} = E_0 (M) + \Delta E_\Upsilon (M,T = 0)
\end{equation}
If non-zero temperature quarkonium states has a pole like
$\Upsilon (1S)$ state \cite{Aarts:2011sm}, 
\begin{equation}
M_\Upsilon (T) = E_0 + \Delta E_\Upsilon (M,T) = M_\Upsilon^{\; \rm exp}
+ \Delta E_\Upsilon (M,T) - \Delta E_\Upsilon (M,T = 0).
\end{equation}
So, the ratio of the non-zero temperature quarkonium correlator to the
$N_t = 80$ correlator can reveal the heavy quark mass dependence
$\Delta E (M,T)$ although the true ground state behaviors for $N_t =
80$ correlators and for $N_t = 16$ or $20$ correlators will not set in
at large $\tau$. Fig. \ref{fig:5} shows such ratios for $T = 458 (N_t
= 16)$ (MeV) and $T = 408 (N_t = 18)$ (MeV). 

In Fig. \ref{fig:5}, the comparison between the filled symbol and the
empty symbol shows the temperature dependence for the same $M$, and
the comparison between the different symbols shows $M-$dependence for
the same $T$. The correlator ratio change due to 10\% change in the
two different $T$ is larger than due to $10 \sim 20 \%$ change in the
4 different $M$. Since the small change in heavy quark mass is less
important than that due to the temperature change, lattice NRQCD is
valid as an effective theory and systematic error due to heavy quark
mass mis-tuning is small.

\section{Conclusion}

Using NRQCD formalism in non-zero temperature, we computed momentum
dependent $(\eta_b, \Upsilon)$ correlators and calculated spectral
functions using MEM via NRQCD kernel. From the peaks and the width of
momentum dependent spectral function, we find that there are
observable $v^2$ effect on the energy of S-wave state moving in
thermal bath but no observable effect on the width of S-wave state
moving in thermal bath for $v^2_{\rm upsilon} \stackrel{<}{\sim} 0.04$.

Among various sources of systematic errors involved in our result,
bottomonium correlators calculation is repeated with $20 \%$ variation
in bottom quark mass. By studying correlator ratios, We find that the
temperature effect is more important than the heavy quark mass effect
in S-wave bottomonium at the temperature around a few $T_c$, which
suggests lattice NRQCD as an effective theory for bottomonium in
non-zero temperature is a consistent theory.

Improved tuning for bottom quark mass and detecting thermal deviations
from the standard dispersion relation is in progress for $N_f = 2 + 1$
flavor system with smaller spatial lattice spacing and a larger
spatial extent.

\section{Acknowledgements}

We acknowledge the support and infrastructure provided by the Trinity
Centre for High Performance Computing and the IITAC project funded by
the HEA under the Program for Research in Third Level Institutes
(PRTLI) co-funded by the Irish Government and the European Union. The
work of CA and GA is carried as part of the UKQCD collaboration and
the DiRAC Facility jointly funded by STFC, the Large Facilities
Capital Fund of BIS and Swansea University. GA and CA are supported
by STFC. SK is grateful to STFC for a Visiting Researcher Grant and
INFN for the visit to Frascati, and is supported by the National
Research Foundation of Korea grant funded by the Korea government
(MEST) No.\ 2011-0026688. SR is supported by the Research Executive
Agency (REA) of the European Union under Grant Agreement number
PITN-GA-2009-238353 (ITN STRONGnet) and the Science Foundation
Ireland, grant no.\ 11-RFP.1-PHY-3201. DKS is supported in part by US
Department of Energy contract DE-AC02-06CH11357. JIS has been
supported by Science Foundation Ireland grant 08-RFP-PHY1462 and
11-RFP.1-PHY-3193-STTF11.

\end{document}